\documentclass[aps,notitlepage]{revtex4}
\usepackage{graphicx}
\usepackage{subfigure}
\begin{document}
\title{Gr\"uneisen parameter studies on heavy fermion quantum criticality}\author{Philipp Gegenwart}
\address{EP VI, Center for Electronic Correlations and Magnetism, Institute of Physics, Augsburg University, 86159 Augsburg, Germany}
\vspace{10pt}

\begin{abstract}
The Gr\"uneisen parameter, experimentally determined from the ratio of thermal expansion to specific heat, quantifies the pressure dependence of characteristic energy scales of matter. It is highly enhanced for Kondo lattice systems, whose properties strongly dependent on the pressure sensitive antiferromagnetic exchange interaction between f- and conduction electrons. In this review, we focus on the divergence of the Gr\"uneisen parameter and its magnetic analogue, the adiabatic magnetocaloric effect, for heavy-fermion metals near quantum critical points. We compare experimental results with current theoretical models, including the effect of strong geometrical frustration. We also discuss the possibility to use materials with divergent magnetic Gr\"uneisen parameter for adiabatic demagnetization cooling to very low temperatures.  
\end{abstract}

%
%
%
\maketitle
%
%

\section{Introduction}

Heavy-fermion (HF) systems are materials with large charge carrier masses, exceeding those of free electrons by 2 or 3 orders of magnitude~\cite{Stewart}. This results in a highly enhanced Sommerfeld coefficient $\gamma=C/T$ at low temperatures~\cite{Andres} and huge effective masses in quantum oscillation experiments~\cite{Reinders}. The observation of superconductivity in CeCu$_2$Si$_2$ with a huge anomaly in the low-temperature specific heat directly demonstrated the itinerant character of heavy fermionic quasiparticles~\cite{Steglich79}. 

The physical properties of these materials at low temperatures are dominated by f electrons and their antiferromagnetic (AF) exchange $J$ with conduction electrons. Similar as for diluted moments in a metallic environment, the Kondo interaction leads to a screening of the local moments by conduction electrons. However, in dense Kondo lattices, the indirect exchange coupling between the moments, by a spin polarization of the conduction electrons, can mediate long-range magnetic ordering. This RKKY interaction grows quadratically with $J$, while the Kondo interaction increases exponentially with $J$. The competition between these two interactions is illustrated in the "Doniach diagram", shown in the left part of Fig. 1. For small $J$, magnetic ordering is found, while beyond a critical $J_c$, the ground state changes to paramagnetic. HF behavior is found in the regime where the two competing interactions are of similar strength. In the last decades numerous examples for magnetically ordered, superconducting or just paramagnetic HF metals have been synthesized and investigated~\cite{Stewart,Stewart01,Flouquetreview,Pfleiderer09}.

\begin{figure}\centering
	\subfigure{\includegraphics[width=0.48\textwidth]{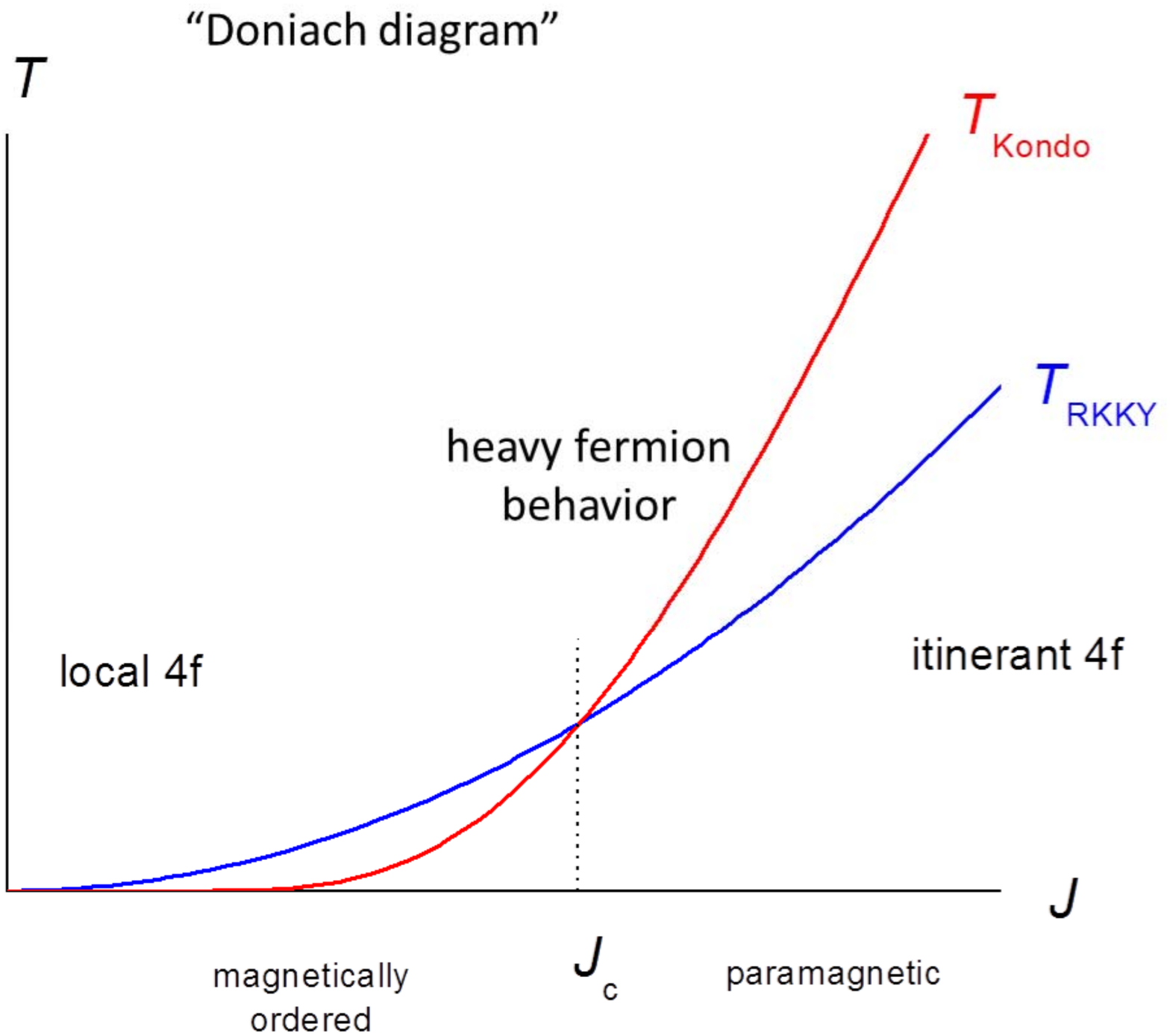}} 
    \subfigure{\includegraphics[width=0.48\textwidth]{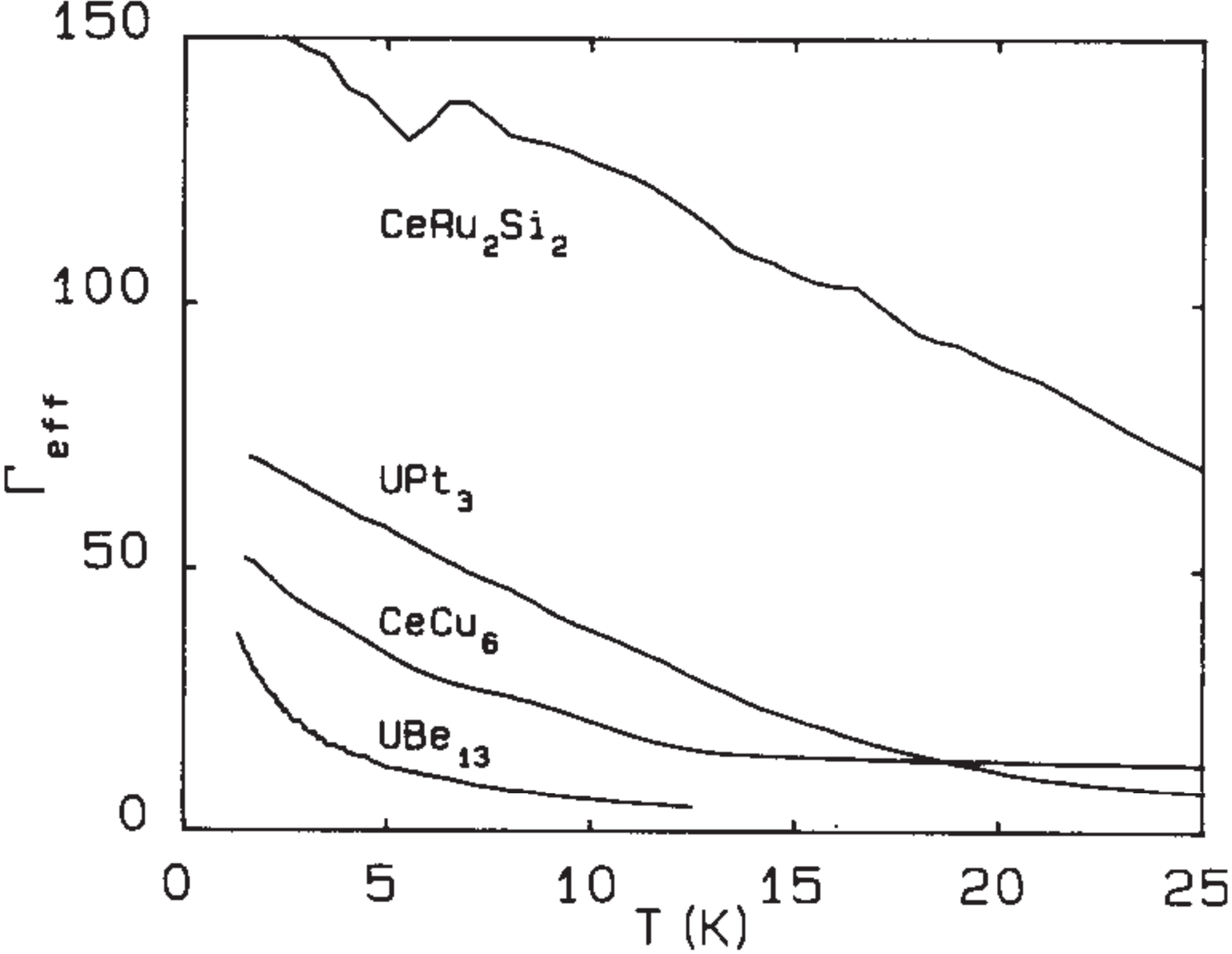}}
    	\caption{Left: Comparison of characteristic energy scales $T_{\rm Kondo}$ and $T_{\rm RKKY}$ versus exchange coupling between f- and conduction-electrons $J$ after Doniach~\cite{Doniach}. Right: Temperature dependence of the "effective" Gr\"uneisen parameter (see text) for different heavy-fermion metals~\cite{DeVisser89}.}
\end{figure}

Indeed, as suggested by Doniach~\cite{Doniach}, HF metals are extremely sensitive to small changes of pressure, which e.g. for Ce-based HF metals enhances $J$ and suppresses magnetic ordering. This pressure sensitivity is reflected in highly enhanced values of the Gr\"uneisen parameter, which we introduce in the following. We start by considering the volume thermal expansion coefficient
\begin{equation}
\beta(T)={1\over V} ({dV\over dT})_p=-{1 \over V} ({dS \over dp})_T,
\end{equation}
i.e., the relative volume change with temperature at constant pressure. Using Maxwell's relation it is proportional to the pressure dependence of the entropy $S$. If entropy has different contributions from e.g. phonons, electrons or magnons, the same holds true for specific heat at constant pressure $C=T(dS/dT)_p$ and thermal expansion. Typically each contribution to entropy can be written as $S(T/T_i)$, where $T_i$ is the respective dominating energy scale. This is the Debye temperature for acoustic phonons, the Fermi energy for a metal or the magnon bandwidth for a magnet. It is then convenient to define Gr\"uneisen parameters $\Gamma_i=V_m/\kappa_T\cdot \beta_i(T)/C_i(T)$ with the molar volume $V_m$ and isothermal compressibility $\kappa_T$.  Inserting the scaling ansatz for the entropy yields
\begin{equation}
 \Gamma_i={\partial T_i/ \partial p\over \kappa_T T_i}=-{\partial \log T_i\over \partial V_m},
  \end{equation}
indicating that the Gr\"uneisen parameters are temperature independent, as found more than 100 years ago by Gr\"uneisen~\cite{Gruneisen}, and that they quantify the relative pressure dependences of the respective energy scales. For phonons in insulators or electrons in simple metals, typically values of order 1 are observed. In the experiment, various contributions to heat capacity and thermal expansion add up and it may be difficult to extract them separately. In this case, one can analyze the "effective Gr\"uneisen parameter"~\cite{DeVisser89}
\begin{equation}
\Gamma_{\rm eff}(T)={V_m\over \kappa_T} {\beta(T) \over C(T)}=\sum \Gamma_i {C_i(T) \over C(T)},
\end{equation}
which is the sum of the Gr\"uneisen parameters from the various contributions times their relative fraction to the total heat capacity. HF metals display a largely enhanced effective Gr\"uneisen parameter (cf. right part of Fig. 1), reflecting that a small volume change has huge influences on these materials~\cite{DeVisser89}. This is because their dominating energy scale, called, Kondo temperature $T_K$, is of order 1-100~K, which is much lower than the Fermi temperature of metals, and highly pressure dependent.

Motivated by the theoretical prediction of a divergent Gr\"uneisen parameter near quantum critical points by Zhu, Garst, Rosch and Si in 2003~\cite{zhu}, more thorough experimental investigations to temperatures below 4 K were performed~\cite{Kuechler03,Kuechler04,Kuechler_Physica06,Kuechler07,Gegenwart10}. Below 4~K the f-electron contribution to heat capacity and thermal expansion largely exceeds the contribution of phonons. Thus, the measured Gr\"uneisen data presented here arise from electronic degrees of freedom and the suffix ``effective'' can be omitted since phonons play hardly any role here.

In the next section, we introduce quantum criticality and discuss its generic signatures in measurements of the Gr\"uneisen parameters. After discussing QCP scenarios (section 3), we first provide a few examples for quantum critical behavior in accordance with the expectations of the Hertz-Millis theory for itinerant quantum criticality (section 4). Section 5 (beyond Hertz-Millis) introduces results which are at odds with those predictions and hint at new forms of QCPs. Section 6 discusses the possibility of cooling through quantum criticality before the paper ends with a summary and outlook in section 7.

\section{Generic signatures of quantum criticality}

\begin{figure}\centering
	\includegraphics[width=\textwidth]{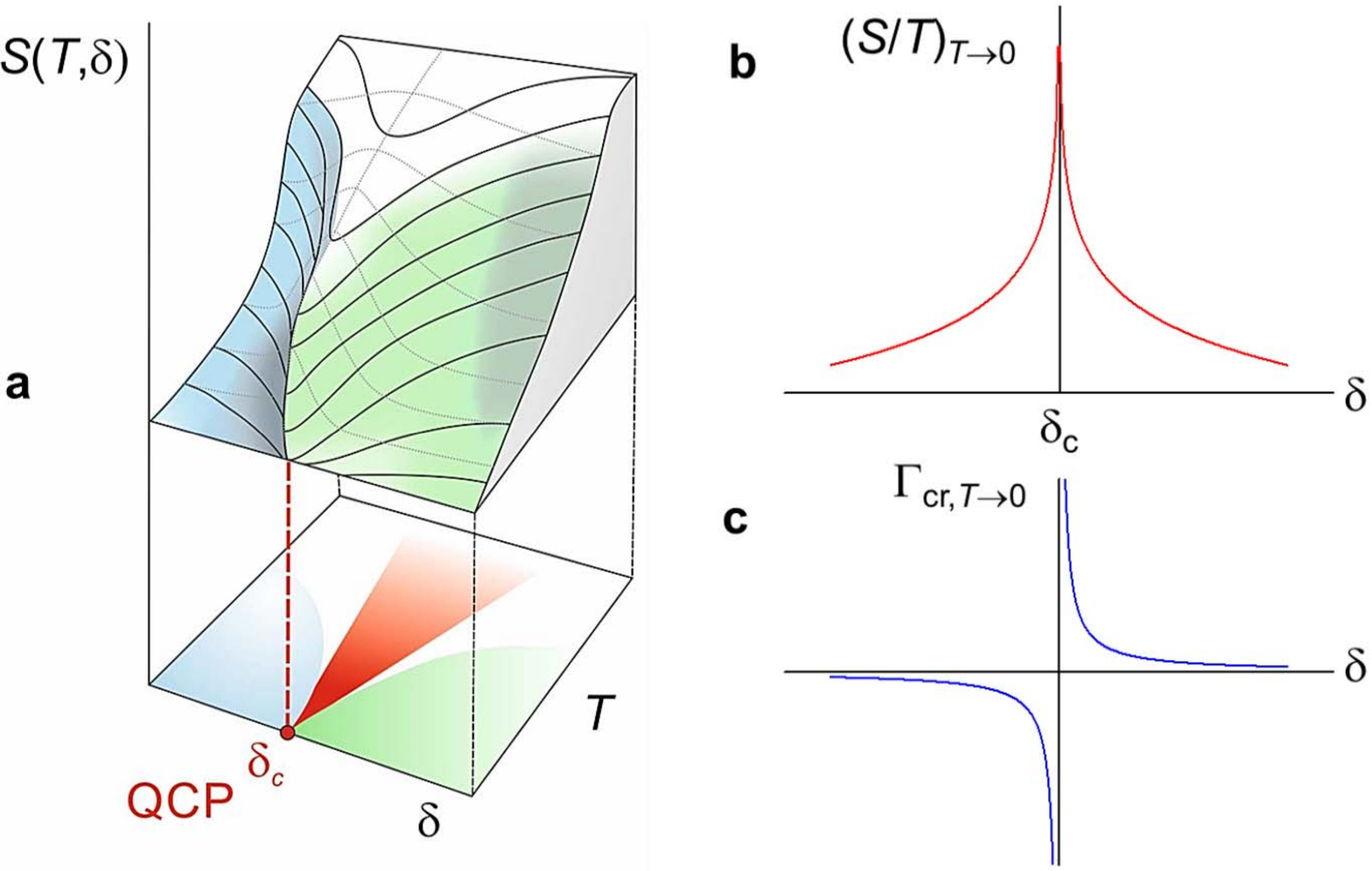}
	\caption{(a) Schematic plot of the entropy ($S$) ridge near a quantum critical point (QCP) at $\delta_c$ where $\delta$ is a non-thermal control parameter, like pressure or magnetic field~\cite{Grube}. The black and grey lines indicate adiabatic temperature ($T(\delta)_S$) and isothermal entropy ($S(\delta)_T$) traces, respectively. (b) and (c) indicate the generic entropy accumulation and divergence of the critical Gr\"uneisen ratio near the QCP~\cite{Wu}.}
\end{figure}

Although absolute zero cannot be reached experimentally, the concept of a continuous phase transition at $T=0$ has been important in wide areas in physics~\cite{Sachdev}. Such QCPs, driven by the variation of a non-thermal parameter, are particularly relevant for strongly correlated electron systems~\cite{Lohneysen,Gegenwart08}. These materials often display a competition of various phases. They realize differing ground states, such as insulating or metallic, paramagnetic or magnetically ordered. For HF metals, as shown in Fig. 1 (left), quantum criticality arises from the competition between the Kondo and the RKKY interaction.

Quantum criticality is qualitatively different from classical criticality. Consider  the energy of temporal order-parameter fluctuations $\hbar/\tau$. As the correlation time $\tau$ diverges in the approach of the critical temperature $T_{\rm c}$ one obtains $\hbar/\tau \ll k_{\rm B}T_{\rm c}$ within the scaling regime close to the phase transition. Consequently temporal order-parameter fluctuations do not play a role for classical criticality. However, $T_{\rm c}=0$ for a QCP. Therefore the above inequality is violated in the critical regime associated with a QCP. In the quantum critical regime the free energy can not be expressed by a function $f(t)$, with $t=(T-T_{\rm c})/T_{\rm c}$, as for classical criticality. It is rather temperature itself, which determines the growth of order parameter fluctuations when the QCP is approached. In the regime above the QCP, the 
energy of order parameter fluctuations is just given by $k_B T$. Therefore 
the (constant) dominating energy scale $T_i$ in equation (2) has to be replaced by temperature and thus a divergence of the Gr\"uneisen parameter upon cooling in the quantum critical regime (cf. the red area in Fig. 2a) is revealed~\cite{zhu}. The quantity $\Gamma(T)$ allows to prove the existence of a QCP. Close to a QCP the correlation length $\xi~\sim |r|^\nu$ ($r=(\delta-\delta_c)/\delta_c$) diverges with an exponent $\nu$. The correlation time $\xi_\tau~\sim\xi^z$ is related to the correlation length by the dynamical critical exponent $z$~\cite{Lohneysen}. Hyperscaling of the free energy reveals, that the ratio of the critical contributions to thermal expansion and heat capacity, called critical Gr\"uneisen ratio, diverges as
\begin{equation}
 \Gamma_{\rm cr}\sim \beta_{\rm cr}/C_{\rm cr}\sim T^{-1/(\nu z)}
  \end{equation}
in the quantum critical regime above the QCP~\cite{zhu}. Furthermore, within the quantum critical regime, the Gr\"uneisen ratio changes its sign upon tuning the control parameter~\cite{Garst05}. This is illustrated in Fig. 2c. At finite temperatures above the QCP,  entropy is enhanced because of the frustration due to the different competing ground states. Respectively, temperature traces at constant entropy (cf. the black lines in Fig. 2a) show a minimum when crossing the quantum critical regime by tuning the control parameter $\delta$.  Since the Gr\"uneisen ratio $\Gamma_{\rm cr}\sim d\ln T/d \delta$ measures the normalized slope of these traces, the entropy accumulation generically results in a sign-change of $\Gamma_{\rm cr}$ within the quantum critical regime. Thus, there are two important characteristics of QCPs: The Gr\"uneisen ratio diverges and changes its sign as function of the control parameter. The absence of an analytical Gr\"uneisen ratio divergence has proven smeared QPTs instead of generic QCPs for nearly ferromagnetic metals Ni$_x$Pd$_{1-x}$~\cite{Kuechler_Physica06} and CePd$_{1-x}$Rh$_x$~\cite{Westerkamp}.

 Determining experimentally the Gr\"uneisen ratio with high precision requires independent measurements of the linear thermal expansion to obtain the volume expansion and measurements of the specific heat. Such experiments can be realized using capacitive dilatometers~\cite{Kuechlerrsi} and micro-calorimeters~\cite{Wilhelm}. In addition the linear thermal expansion 
can reveal important information on anisotropic behavior and lattice symmetry breaking transitions~\cite{stingl11}.

In many cases, QCPs can be tuned by magnetic fields. It is therefore best to study the "magnetic" Gr\"uneisen parameter which is thermodynamically given by the ratio of the negative temperature derivative of the magnetization (which equals the field derivative of the entropy) to the heat capacity:
\begin{equation}
 \Gamma_H={-(dM/dT)\over C}=T^{-1} ({\partial  T \over \partial H})_S
  \end{equation}
Interestingly, this property equals the adiabatic magnetocaloric (MCE) effect. In contrast to the ordinary Gr\"uneisen parameter, the magnetic Gr\"uneisen parameter can be measured directly by determining the temperature change induced by a field change under adiabatic conditions. High-precision measurements of the adiabatic MCE down to very low temperatures have been realized utilizing an alternating magnetic field method. The so-derived adiabatic MCE was confirmed to exactly equal the magnetic Gr\"uneisen ratio calculated  from independent measurements of the magnetization and specific heat~\cite{Tokiwarsi}. In contrast to pressure, it is easy to change the magnetic field in-situ at low temperatures. This makes the adiabatic MCE an extremely useful probe for the investigation of QCPs.

\section{Quantum critical point scenarios for Heavy-fermion systems}

\begin{figure}\centering
	\includegraphics[width=\textwidth]{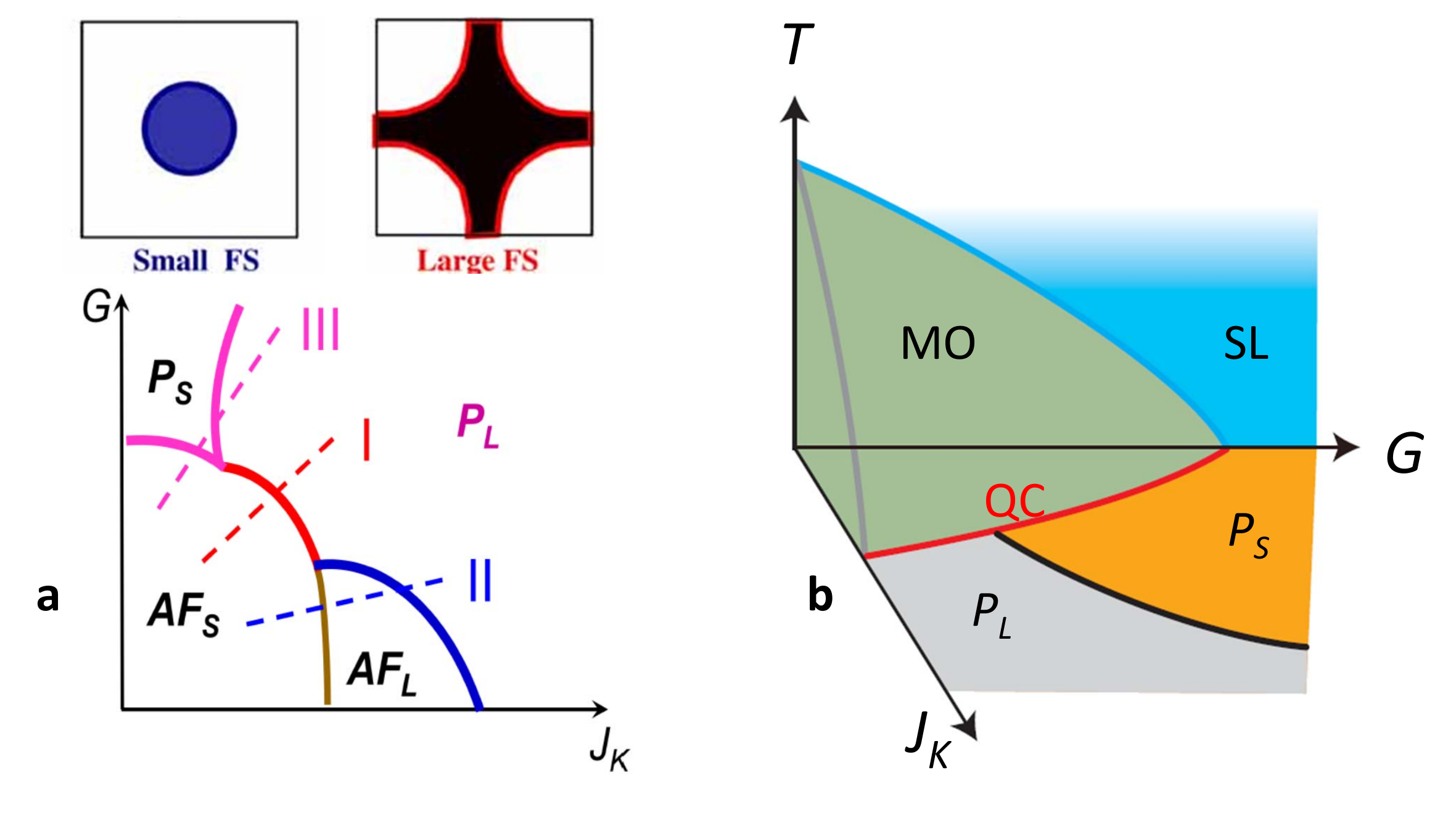}
	\caption{Schematic "global phase diagram" proposed for AF Kondo lattices at $T=0$~\cite{Si10} (a). $G$ represents the strength of quantum fluctuations resulting from magnetic frustration or low spatial dimensionality, $J_K$ denotes the normalized Kondo coupling. Ground states are distinguished with respect to their magnetic (P: paramagnetic, AF: antiferromagnetic) and electronic (subscripts "S" and "F" denote small and large Fermi surface volume, as sketched, respectively) characteristics. The three dashed lines indicate possible trajectories across different quantum phase transitions. (b): Three-dimensional $T$-$G$-$J_K$ phase diagram (schematic sketch). The blue sector in the $J_K=0$ plane indicates an insulating spin liquid (SL). A line of quantum critical points (in red) bounds
the long-range magnetic order (MO) (green region) at $T=0$~\cite{Tokiwa14}.}
\end{figure}

As discussed above quantum criticality in HF metals arises from the competition between the on-site Kondo screening of f-moments and the AF exchange coupling between the moments~\cite{Gegenwart08}. In the paramagnetic (P) regime, where the Kondo interaction dominates, a heavy Fermi liquid (FL) ground state is expected. The heavy charge carriers are considered as composite bound states formed between the local moments and the conduction electrons. The central question is what happens to these Kondo singlets as a material is tuned by variation of some non-thermal control parameter to a magnetic QCP~\cite{Coleman,Si}. In the Kondo screened state, the Fermi surface (FS) volume counts both the number of conduction electrons and number of bound states and is therefore called "large". On the other hand, magnetic ordering could either be of itinerant, i.e. spin density wave (SDW), or local moment character and respectively the FS volume in the ordered state is either large or small, respectively. For a small FS volume, the Kondo singlet formation must be destructed; strong quantum fluctuations have been proposed as possible mechanism acting against the Kondo effect~\cite{Si10,Vojta08}.

In Fig. 3a~\cite{Si10}, the strength of quantum fluctuations is parametrized by the symbol $G$. Large values of $G$ result from a reduced dimensionality of magnetic interactions and/or strongly frustrated interactions. Fig. 3a represents the "global phase diagram" for the ground state of Kondo lattice metals~\cite{Si10}. Its four different ground states are arising from $2\times 2$ possibilities for the magnetic (P or AF) and electronic (large of small FS) degrees of freedom.  To experimentally settle the magnetic ground state (ordered or disordered) is straightforward. However, it is harder to probe the FS volume. This could be done using quantum oscillation measurements. However their observation for HF metals requires very clean single crystals (typically excluding the possibility for fine-tuning a material near a QCP by chemical substitution) and the combination of very low temperatures and very large magnetic fields, due to the large quasiparticle masses. The last restriction could be problematic if large fields are suppressing the quantum critical fluctuations and polarizing the Kondo singlets. Angular resolved photoemission spectroscopy (ARPES) on the other hand does not require a magnetic field and can be observed on doped samples as well. However, it is surface dependent and obtaining data at the required very low temperatures with necessary energy resolution (related to low characteristic energy scales below 1~meV) is also extremely ambitious. 
The dashed lines in Fig.~3a indicate possible trajectories through QCPs. For instance, type I has a "locally critical" QCP~\cite{Si} where the f-electrons localize due to a local destruction of Kondo singlet formation, while the FS volume does not change at the P$_L$ to AF$_L$ transition along trajectory II. Comparing experimental results with these trajectories~\cite{Custers,Paschen} is difficult because it is not clear how $G$ and $J_K$ change under the variation of pressure, composition or magnetic field tuning. Note, that neither $G$ nor $J_K$ can be quantified experimentally for Kondo lattices. A jump of the Hall constant $R_H$ has been proposed as indication for a locally critical QCP~\cite{Coleman}. For YbRh$_2$Si$_2$ a crossover of $R_H$ near an energy scale $T^\ast(B)$ which extrapolates for $T\rightarrow 0$ to the QCP has been found, whose width obeys a linear temperature dependence~\cite{Paschen04,Friedemann10}. Under the assumptions the Hall crossover anomaly is a finite temperature signature of a FS change and that the extrapolation of its width to $T=0$ indicates jump of the FS volume, the data would be compatible with a locally critical QCP of trajectory I (see later).

The three-dimensional sketch in Fig. 3b indicates how finite temperature magnetic ordering can be weakened by increasing either the Kondo coupling or the strength of quantum fluctuations~\cite{Tokiwa14}. For magnetic insulators with $J_K=0$, $G$ can be quantified by the ratio between the average size of the magnetic coupling to the ordering temperature (called "frustration parameter"). Increasing $G$ leads to a spin liquid (SL) regime above the ordering temperature, which eventually extends to $T=0$. Theoretically it has been proposed, that for sufficiently large 
frustration in Kondo lattices, the f-electrons are decoupled from the conduction electrons. Within the P$_S$ regime, a so-called fractionalized Fermi liquid has been proposed which is an exotic metallic spin liquid phase~\cite{Vojta08}, indicated by orange color in Fig. 3b. This proposal motivates to investigate highly geometrically frustrated Kondo lattices (see below) which should lead to a connection between the research areas of spin liquid physics in insulators and quantum criticality in Kondo lattices~\cite{Tokiwa14}.

\section{Hertz-Millis type behavior}

\begin{figure}\centering
	\includegraphics[width=\textwidth]{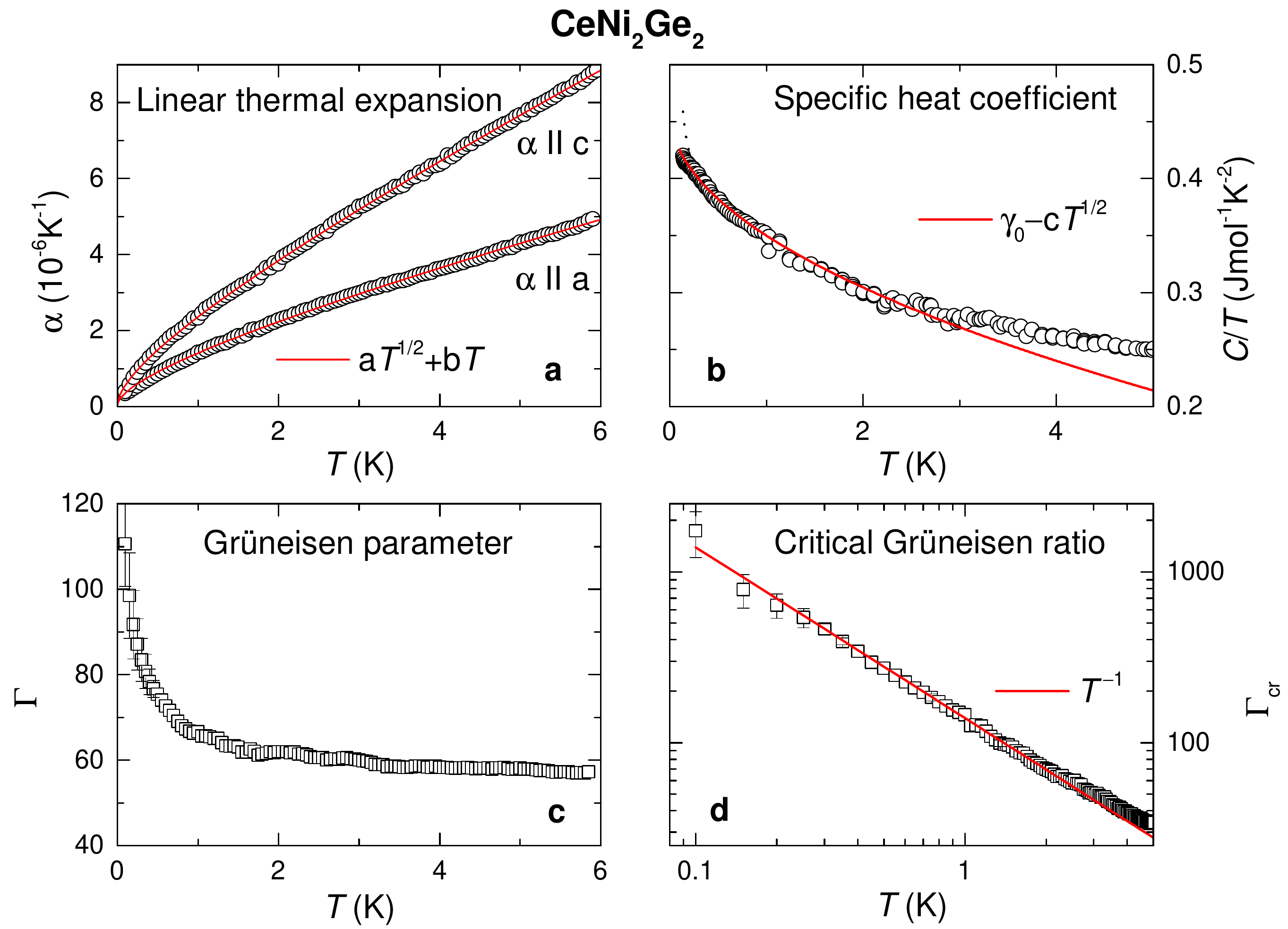}
	\caption{Low-temperature thermodynamic properties of the paramagnetic Kondo lattice CeNi$_2$Ge$_2$~\cite{Kuechler03}. Temperature dependence of the linear thermal expansion (a), specific heat coefficient (b), Gr\"uneisen parameter (c) and critical Gr\"uneisen ratio (d). Red solid lines indicate quantum criticality in accordance with predictions of the Hertz-Millis theory for 3D AF critical fluctuations. The dotted line in (b) indicates the raw data, including a nuclear contribution.}
\end{figure}

As discussed above an important question in the context of HF physics concerns a possible change of the FS volume at the QCP. If there is no such change the itinerant theory for quantum criticality in metals should be applicable. The description of quantum critical order parameter fluctuations in metallic environment has first been developed by Hertz~\cite{Hertz} and Millis~\cite{Millis}. Here the electronic degrees of freedom are "integrated out" and criticality is determined by overdamped bosonic modes which are analyzed by renormalization group methods. The Hertz-Millis scenario considers the spatial and temporal order parameter fluctuations in the presence of Landau damping by charge carriers, leading to $\nu=1/2$ and a dynamical critical exponent $z=2$ or 3 for AF or ferromagnetic (FM) critical fluctuations, respectively~\cite{Lohneysen}. For AF quantum criticality in three spatial dimensions ($d=3$), critical fluctuations in reciprocal space are confined to thermally smeared regions near hot spots at the FS which are connected by the wave vector of nearby SDW ordering. Consequently, large portions of the FS are noncritical giving rise to FL contributions to
thermal expansion and heat capacity~\cite{zhu}.

For thermal expansion and specific heat, this results in the temperature dependences indicated by the red lines in Fig. 4a and b. The data for the nearly AF HF metal CeNi$_2$Ge$_2$ indicate a zero-field AF QCP of Hertz-Millis type~\cite{Gegenwart99,Kuechler03}. The low-temperature specific heat coefficient follows $C/T=\gamma_0-aT^{1/2}$, where the critical square-root term is sub-leading at low temperatures~\cite{zhu}. This is different in the case of thermal expansion. Here quantum criticality leads to a true divergence of the coefficient to $T\rightarrow 0$ with a  singular $T^{-1/2}$ contribution in $\alpha/T$. Respectively, the Gr\"uneisen parameter, derived from the total thermal expansion and specific heat, diverges upon cooling (Fig. 4c). For comparison with the theoretical prediction, the critical Gr\"uneisen parameter is calculated from the ratio of the critical contributions to thermal expansion and heat capacity. As shown in Fig. 4d, it diverges as $T^{-1}$ for CeNi$_2$Ge$_2$. Using equation (4), this is in perfect agreement with the scaling prediction ($\nu=1/2$ and $z=2$) expected from the itinerant Hertz-Millis theory~\cite{Kuechler03}. 

\begin{figure}\centering
	\includegraphics[width=\textwidth]{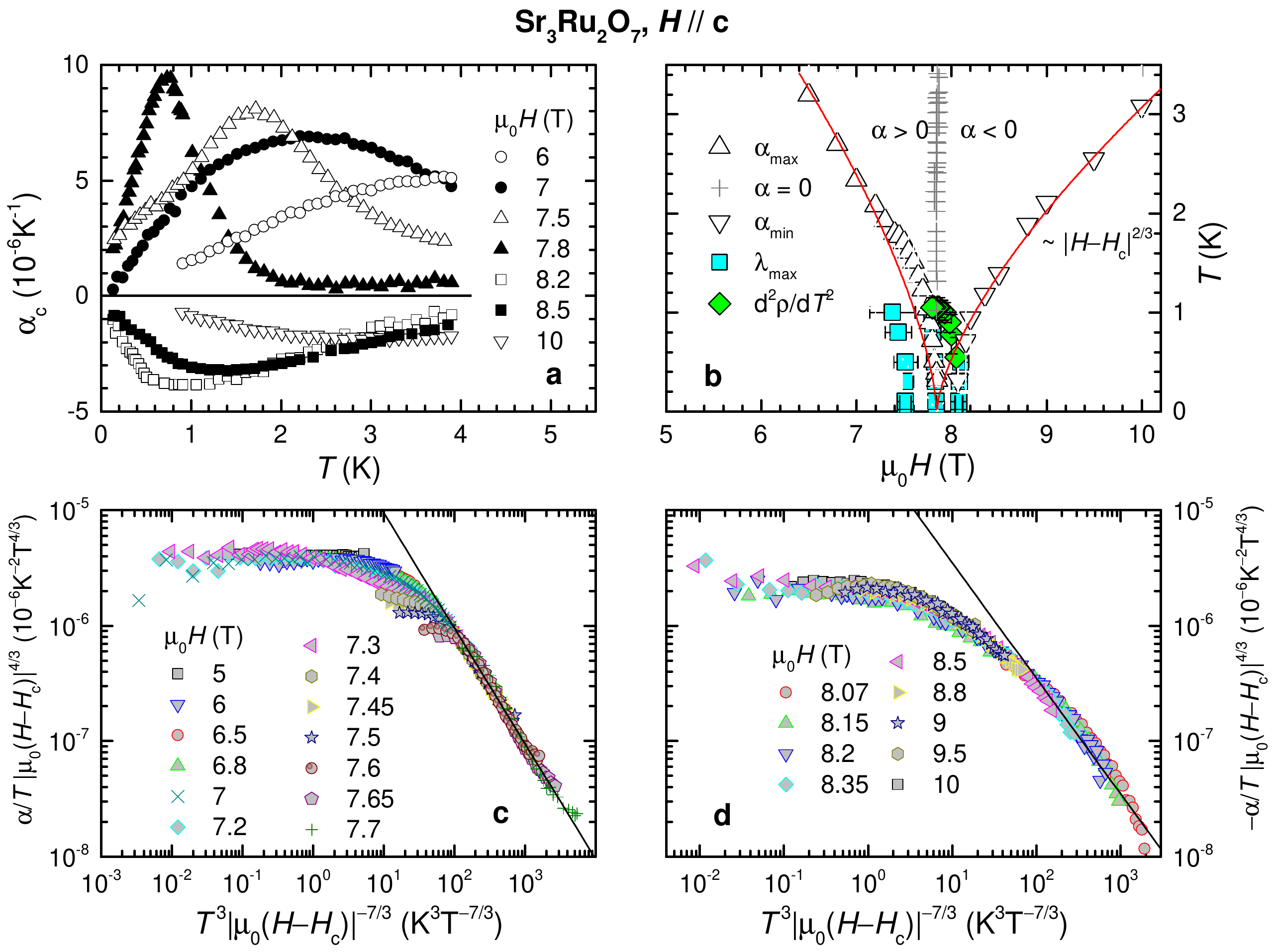}
	\caption{Thermal expansion study of metamagnetic quantum criticality in Sr$_3$Ru$_2$O$_7$ for fields $H\parallel c$~\cite{GegenwartSR327}. Temperature dependence of the linear $c$-axis thermal expansion coefficient (a), temperature-magnetic field phase diagram (b) and quantum critical scaling according to the Hertz-Millis theory for a 2D quantum critical end point for $H<H_c$ (c) and $H>H_c$ (d) with critical field $\mu_0 H_c=7.845$~T.}
\end{figure}

Within the class of AF Kondo lattice materials, quantum criticality in accordance with this theory has also been found for cubic CeIn$_{3-x}$Sn$_x$ near $x_c=0.65$~\cite{Kuechler06} and tetragonal  CeRhIn$_{5-x}$Sn$_x$ near $x_c=0.48$~\cite{Donath09}.

Hertz-Millis type quantum criticality has also been expected near a quantum critical end point (QCEP), related to itinerant metamagnetism~\cite{Millis02}. Here, the theoretical description is based upon longitudinal fluctuations of the magnetization density with a dynamical critical exponent $z=3$. To realize such a scenario, a line of first-order itinerant metamagnetic transitions ending at a finite-temperature critical point is suppressed to $T=0$ by variation of an additional non-thermal parameter (e.g. pressure). The bilayer strontium ruthenate Sr$_3$Ru$_2$O$_7$ has been discussed in this context~\cite{Grigera01}. Indeed, there are strong indications from electrical resistivity and specific heat for a field-induced QCP near 8 T for $H\parallel c$~\cite{Grigera01,Rost}. However, in the vicinity of the putative QCP below 1~K a symmetry-broken phase was found~\cite{stingl11,Grigera04}, shown in the temperature-field diagram of Fig. 5b. This phase has recently been identified as SDW ordering~\cite{Lester}. Importantly, above this phase, thermal expansion is in full agreement with the predictions of the itinerant metamagnetic QCEP scenario~\cite{GegenwartSR327}. Most recently, another metamagnetic QCEP at 7.5~T  has been established and the interference between between both instabilities has been clarified~\cite{Tokiwa16}.

Fig. 5a displays the temperature dependence of the c-axis thermal expansion at fields below and above the critical field, whose value has been precisely determined from the scaling analysis as $H_c=7.845$~T. Characteristic maxima and minima are observed, whose positions, cf. the triangles in Fig. 5b, extrapolate to the QCEP. Furthermore, $\alpha$ displays a sign change at finite temperatures above the critical field. Using the relation $\alpha=-V_m^{-1}(dS/dp)=\Omega/V_m(dS/dH)$ with the constant $\Omega=(dH_c/dP)$ it is evident, that these zero-crossings of $\alpha(H)$ indicate the entropy accumulation above the QCEP, while the thermal expansion extrema are related to the crossover between the quantum critical and FL regimes in the phase diagram~\cite{GegenwartSR327}. Besides these qualitative signatures of a field-tuned QCP, the data perfectly fit the predictions of the Hertz-Millis theory for a two-dimensional QCEP ($z=3$)~\cite{GegenwartSR327}. This is demonstrated in the two scaling plots (Fig. 5c and d) for fields below and above the critical field.

For a more comprehensive discussion of the universal thermodynamic signatures at a metamagnetic QCEP and their observation in the HF metal CeRu$_2$Si$_2$, we refer to Weickert {\it et al.}~\cite{Weickert}. Within the scaling regime all second-order derivatives of the free energy display a similar divergence upon approaching the critical field, resulting in a proportionality of differential susceptibility, magnetostriction and electronic compressibility. This generates divergences of all these properties. A divergence of the electronic compressibility may lead to a significant softening of the crystal lattice and may even give rise to structural instabilities~\cite{Zacharias}.
The above mentioned CeRu$_2$Si$_2$ is actually located slightly off a QCEP and the quantum critical regime is at any field confined to temperatures above 0.5~K~\cite{Weickert}. Nevertheless, an approximately 50\% reduction of the elastic modulus was found near the critical field~\cite{Flouquet}. In such cases, it may become important to consider the feedback of the lattice to quantum critical behavior~\cite{Zacharias}. A generic metamagnetic QCEP could actually be intrinsically unstable and preempted by a structural quantum phase transition. Similar behavior is expected at valence QCEPs.

\section{Beyond Hertz-Millis}

\begin{figure}\centering
	\includegraphics[width=\textwidth]{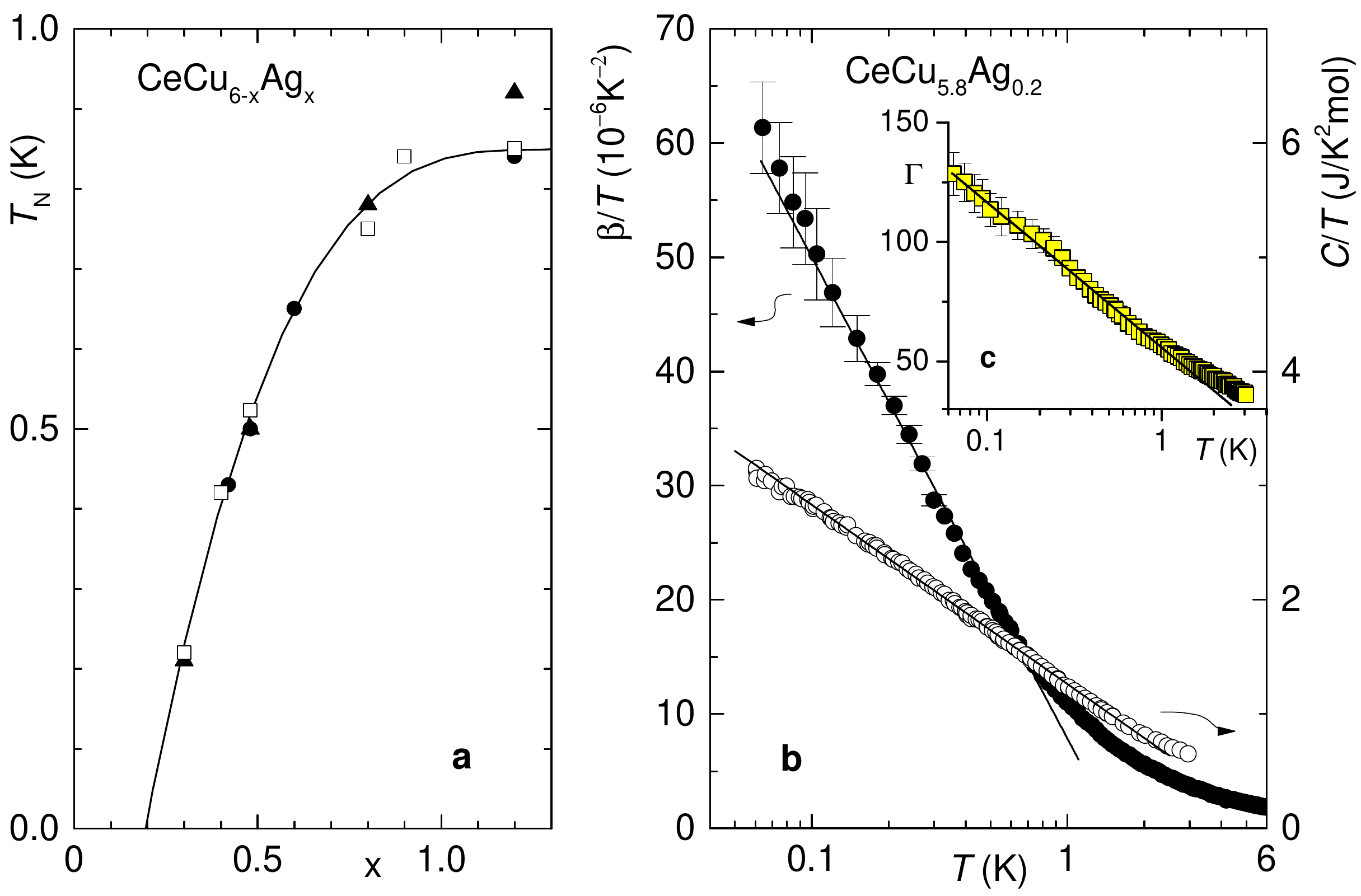}
	\caption{Quantum criticality in CeCu$_{6-x}$Ag$_x$~\cite{Fraunberger,Kuechler04}. (a) AF ordering temperature $T_N$ vs. Ag concentration $x$. (b) Temperature dependence of the  volume thermal expansion coefficient (left axis) and specific heat coefficient (right axis) for quantum critical CeCu$_{5.8}$Ag$_{0.2}$. Inset (c) Gr\"uneisen parameter $\Gamma(T)$. Lines in (b) and (c) indicate logarithmic temperature dependences.}
\end{figure}

In the following, we will discuss experimental results for which the Hertz-Millis description of quantum criticality fails. First we note, that such failure must not immediately indicate a FS change due to a Kondo destruction. While the Hertz-Millis-description by noninteracting Gaussian fluctuations (weak coupling scenario) predicts that critical fluctuations are centered near "hot lines" of the Fermi surface, separated by the critical wave-vector $Q$ of the long-range ordering~\cite{Hertz,Millis}, a strong coupling scenario leading to critical quasiparticles at the entire FS and a mass divergence has recently been proposed, which does not require a FS change~\cite{Abrahams}.

 We first concentrate on doped CeCu$_6$ and YbRh$_2$Si$_2$. Subsequently, we will discuss new types of quantum criticality arising in the geometrically frustrated Kondo lattices YbAgGe and CeRhSn.

For CeCu$_{6-x}$Au$_x$ the partial isoelectronic substitution of Cu by Au induces a negative chemical pressure and leads to pronounced NFL behavior near a magnetic instability at $x_c=0.1$~\cite{Lohneysen94}. Surprisingly, the critical magnetic fluctuations in this material near the QCP have a quasi-two-dimensional character~\cite{Stockert}. Even more important, the dynamical susceptibility, extracted from inelastic neutron scattering, displays an anomalous energy over temperature scaling, incompatible with the Hertz-Millis-theory~\cite{Schroeder}. Remarkably, such scaling is found not only near the critical wave vector of the nearby AF ordering but rather over the entire Brillouin zone. This wave-vector independence led to the proposal of local quantum criticality due to a destruction of the Kondo effect~\cite{Si}. Indeed, this scenario reproduces the unusual fractional scaling exponent in doped CeCu$_{6-x}$Au$_x$.

The low-temperature specific heat coefficient displays a logarithmic divergence near the QCP in this material, independent on whether the instability is reached by chemical substitution only or by combination of hydrostatic pressure and substitution~\cite{Bogenberger}. Similar behavior is also found in the related series CeCu$_{6-x}$Ag$_x$ near its QCP at $x_c=0.2$, cf. Fig. 6a~\cite{Fraunberger,Kuechler04}. A study of the thermal expansion and Gr\"uneisen parameter has confirmed that also thermodynamic properties are incompatible with the Hertz-Millis theory: The logarithmic divergence of the specific-heat coefficient would in this theory require 2D AF spinfluctuations. However, for this case a leading $1/T$ dependence of the thermal expansion coefficient and of the Gr\"uneisen ratio would have been predicted. This is clearly in contrast with the experimental data (Fig. 6b and c), which display a logarithmic divergence of the two properties only. It has recently been suggested that the fluctuations in a generic AF model for itinerant fermions could be mapped to those in the universality class of the dissipative quantum-XY model~\cite{ZhuChenVarma}. In the 2D case, this consistently explains the observed energy over temperature scaling in CeCu$_{5.9}$Au$_{0.1}$~\cite{Varma15}. In the same scenario, the correlation length depends logarithmically on the correlation time. For thermodynamic properties such as the Gr\"uneisen ratio this implies results similar as obtained by taking $z=\infty$, yielding $\Gamma\propto -\ln T$~\cite{Varma} in full accordance with the experimental observations.

\begin{figure}\centering
	\includegraphics[width=\textwidth]{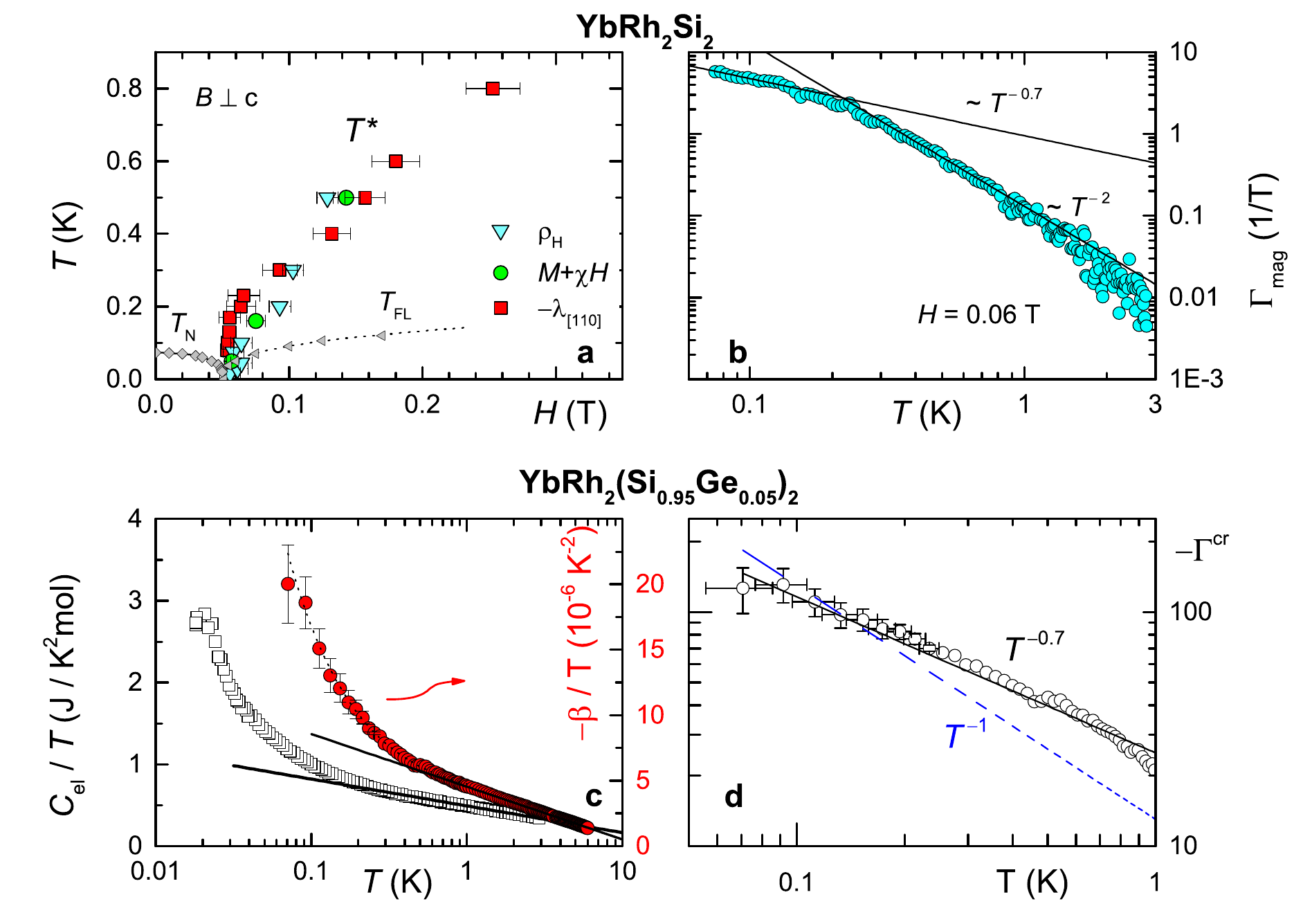}
	\caption{Quantum criticality in YbRh$_2$(Si$_{1-x}$Ge$_x$)$_2$. (a) Temperature--magnetic field phase diagram for $x=0$ where $T_N$, $T^\ast$ and $T_{\rm FL}$ denote the AF ordering temperature, an additional low-energy scale (see text) and the upper boundary of $T^2$ behavior in electrical resistivity, indicative of FL behavior, respectively~\cite{Gegenwart07}. (b)~Temperature dependence of the magnetic Gr\"uneisen ratio for $x=0$ at the critical magnetic field~\cite{Tokiwa09}. (c)~Specific heat and volume thermal expansion as $C/T$ (left axis) and $-\beta/T$ (right axis) for $x=0.05$ vs. Temperature (on log scale)~\cite{Kuechler03}. (d) Temperature dependence of the Gr\"uneisen ratio for $x=0.05$ on double log scales.}
\end{figure}

Tetragonal YbRh$_2$Si$_2$ belongs to the most intensively studied quantum critical HF metals~\cite{Gegenwart08,trov00}. It displays a very weak AF ordering at $T_N=70$~mK, which is suppressed by small critical magnetic fields $H_c$ which equal 0.06 T ($H\perp c$) and 0.66 T ($H\parallel c$)~\cite{Gegenwart02}. The temperature-field phase diagram of the material is displayed in Fig. 7a. Beyond the QCP at $H>H_c$, a FL ground state develops below a crossover temperature $T_{\rm FL}$, which increases with increasing magnetic field. A detailed analysis of the heat capacity, electrical resistivity and magnetic susceptibility within the FL regime indicates that the quasiparticle mass diverges in the approach of the QCP, which is incompatible with the Hertz-Millis theory for 3D AF spinfluctuations~\cite{Gegenwart02,Custers,Gegenwart05}. Fig.~7b shows the temperature dependence of the magnetic Gr\"uneisen ratio at the critical magnetic field, i.e., upon approaching the QCP by cooling~\cite{Tokiwa09}. Interestingly, a crossover between two different temperature exponents is found near 0.3~K. Neither the high- nor the low-temperature behavior is compatible with the expected $1/T$ divergence. The same also holds for the Gr\"uneisen ratio of thermal expansion to specific heat, determined by slightly Ge-doped YbRh$_2$Si$_2$ (cf. Fig. 7c and d), for which the critical field is very close to zero (only about 0.02 mT). Again distinct crossovers near 0.3~K are observed~\cite{Kuechler03}.

As indicated by the colored symbols in Fig. 7a, the low-temperature phase diagram of YbRh$_2$Si$_2$ contains an additional energy scale, called $T^\ast(H)$ which terminates at the critical field~\cite{Gegenwart07}. Upon crossing this scale, distinct crossovers are found in the isothermal magnetoresistance, Hall effect~\cite{Paschen}, magnetostriction, magnetization and entropy~\cite{Tokiwa09}. Importantly the full width at half maximum of these crossovers displays a linear temperature dependence. Extrapolation to $T=0$ thus suggests a discontinuous change of these properties at the QCP~\cite{Friedemann10}. For the Hall crossover, this would imply a jump of the FS volume~\cite{Paschen04}, as expected in a Kondo breakdown scenario~\cite{Si}. While the AF phase boundary is highly sensitive to pressure, only a very weak pressure dependence of $T^\ast(H)$ was found~\cite{toki09b}. The same holds for positive and negative chemical pressure, induced by partial Co an Ir substitution of Rh, respectively~\cite{frie09}. For the different compositions $T^\ast(H)$ always terminates for $T=0$ at a finite field of about 0.06~T and obeys an approximately a linear field dependence above 0.3~K (cf. Fig. 7a). An explanation could be that $T^\ast$ is induced by magnetic field and not related to a Kondo breakdown. A simple Zeeman-driven Lifshitz scenario has been proposed and criticized as unphysical~\cite{Hackl,Hackl-comment}. More recently $T^\ast$ has been associated with critical quasiparticles undergoing spin-flip scattering off collective spin excitations that is controlled by Zeeman splitting~\cite{WA}.
The observed fractional exponent (0.7) of the low-$T$ Gr\"uneisen exponents (cf. Fig. 7b and d) would be consistent with either the locally-critical QCP scenario~\cite{Kuechler03} or the theory of critical quasiparticles~\cite{Abrahams,WA}. Within the latter, 3D AF quantum criticality with scaling parameters $\nu= 1/3$ and $z=4$ has been assumed~\cite{Wolfle}.

The global phase diagram (Fig. 3a) suggests new kinds of quantum phase transitions for geometrically frustrated Kondo latices~\cite{Si10,Vojta08}. Since strong frustration opposes Kondo singlet formation, a locally critical QCP and even a metallic spin liquid state with local moments is expected for highly frustrated materials. We therefore now turn our attention to quantum criticality in HF systems such as CePdAl~\cite{Fritsch}, YbAgGe~\cite{Canfield} and CeRhSn~\cite{Kim} that crystallize in the hexagonal ZrNiAl structure. In these materials, the f moments are located on equilateral corner-sharing triangles in the $ab$ plane. Such a distorted Kagome lattice is sketched in the inset of Fig. 9b. CePdAl shows a ``partially frustrated'' AF ground state. It has been found by powder neutron scattering, that below $T_{\bf N}=2.7$~K one third of the f moments do not participate in long range order~\cite{Donni}. This may be related to frustration arising from next and second-neighbor in-plane exchange. It is very interesting to study quantum criticality in geometrically frustrated Kondo lattice systems~\cite{Burdin}. Strong quantum fluctuations, induced by frustration, are expected to counteract Kondo singlet formation~\cite{Senthil,Vojta08,Si,Si10}. New kinds of quantum criticality beyond the itinerant Hertz-Millis scenario may therefore be realized for strongly frustrated materials~\cite{Coleman2010,Custers2010,Aronson}. In the highly frustrated limit a "fractionalized" FL is predicted. It consists of local moments in a spin liquid state that are decoupled from the conduction electrons~\cite{Senthil,Vojta08}. For CePdAl negative chemical pressure induces a QCP in CePd$_{1-x}$Ni$_x$Al near $x_c=0.144$~\cite{Fritsch}. Interestingly, the specific heat coefficient at $x_c$ displays a logarithmic divergence similar as found for CeCu$_{5.9}$Au$_{0.1}$. It will be interesting to investigate the critical spinfluctuations by inelastic neutron scattering in this material.
Below we discuss recent results the isostructural hexagonal HF metals, YbAgGe and CeRhSn, with respective f moments on the distorted Kagome configuration, for which the Gr\"uneisen analysis has been performed.

\begin{figure}\centering
	\includegraphics[width=\textwidth]{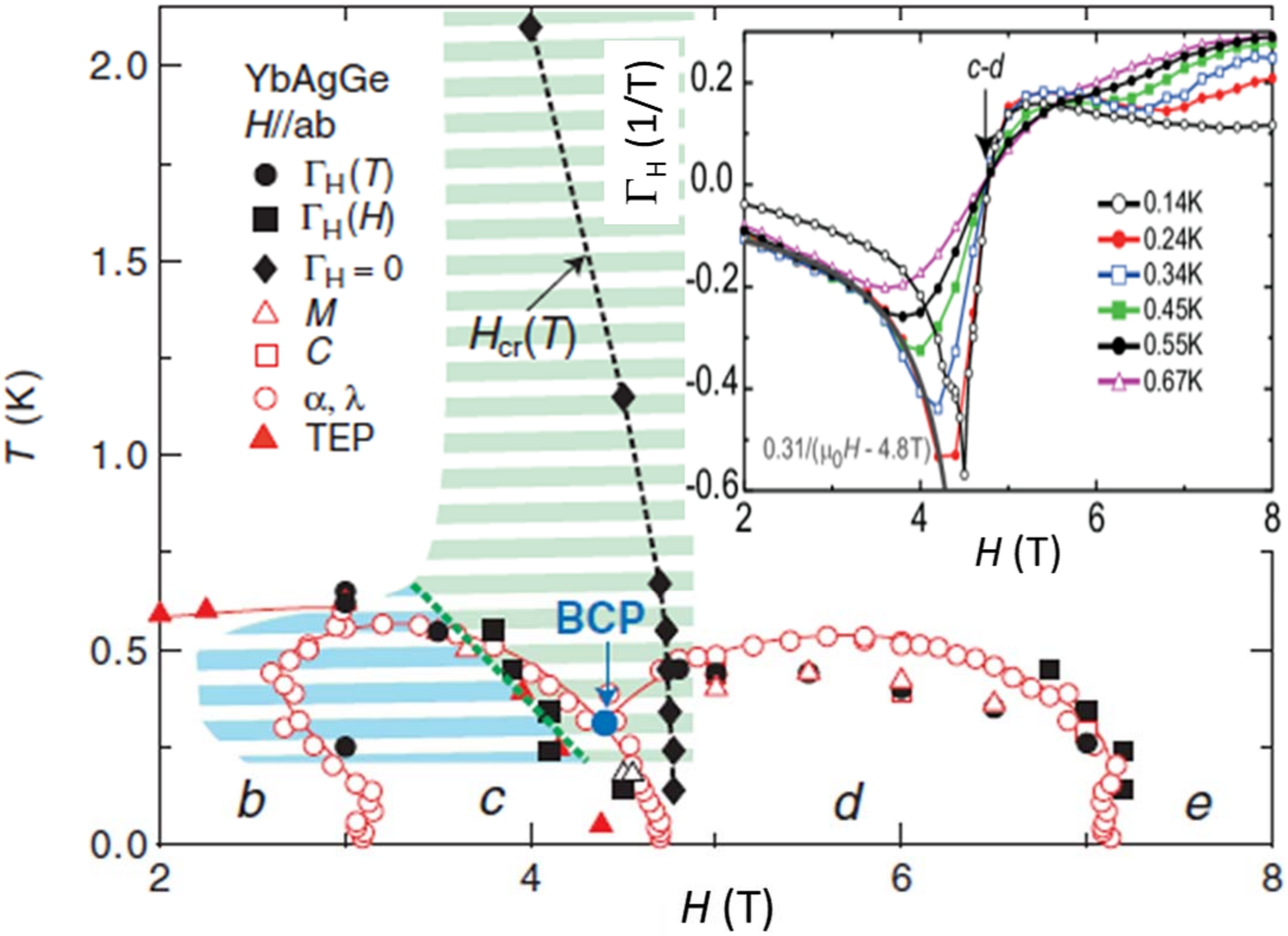}
	\caption{Low-temperature magnetic field phase diagram of YbAgGe as derived from various thermodynamic experiments for magnetic fields between 2 and 8 T (\cite{Schmiedeshoff,TokiwaYbAgGe} and references therein). The blue arrow indicates the position of a bicritical point. Within the shaded regimes, the magnetic Gr\"uneisen parameter (cf. Inset) displays quantum bicritical scaling behavior~\cite{TokiwaYbAgGe}.}
\end{figure}

For YbAgGe the effect of magnetic frustration is evident from the presence of various almost degenerate magnetic states below 1 K which are arising when a magnetic field is applied within the hexagonal planes~\cite{Schmiedeshoff,TokiwaYbAgGe}. Previous transport and thermodynamic signatures have suggested a field-induced QCP near 4.5~T~\cite{Dong}. As shown in the inset of Fig. 8, the low-temperature isothermal magnetic Gr\"uneisen ratio between 2 and 5 T indeed displays the main characteristics for a QCP, which are a diverging behavior and a sign change at 4.8~T (indicating the entropy accumulation point), which coincides with the $T=0$ transition field between phases c and d. However, the boundaries of these phases actually merge at a bicritical point (BCP) at $T_{\rm BCP}=0.3$~K (and $H=4.5$~T, cf. the blue point in Fig. 8), below which the c-d transition is weakly first order~\cite{TokiwaYbAgGe}. The global phase diagram of Fig. 3a neither explicitly treats magnetic field as tuning parameter nor considers the possibility of a metamagnetic (spin flop) transition in the AF$_S$ regime. As sketched in Fig. 9d, a bicritical point as found in YbAgGe arises naturally between different AF$_S$ states with local moments as function of field. Geometrical frustration, which enhances the degree of quantum fluctuations, is efficient in depressing $T_{\rm BCP}$ to absolute zero. Although YbAgGe does not exactly realize such a quantum BCP, $T_{\rm BCP}$ is low enough to lead to pronounced quantum bicritical fluctuations, which dominate the physical properties of this material over large areas in phase space (cf. the shaded regimes of Fig. 8). Indeed the magnetic Gr\"uneisen data can be scaled as $h\Gamma_H$ vs. $T/|h|^{1.1}$~\cite{TokiwaYbAgGe}. This is {\it quantum} in contrast to classical scaling, which would have required a finite critical temperature. A characteristic feature of such quantum bicriticality is the observed asymmetry of the Gr\"uneisen ratio for fields below and above the critical field (inset Fig. 8). Interestingly, the analysis of the scaling exponents suggests low-dimensional, quasi 1D critical fluctuations, which may be promoted by the geometrical frustration in this system~\cite{TokiwaYbAgGe}.

\begin{figure}\centering
	\includegraphics[width=\textwidth]{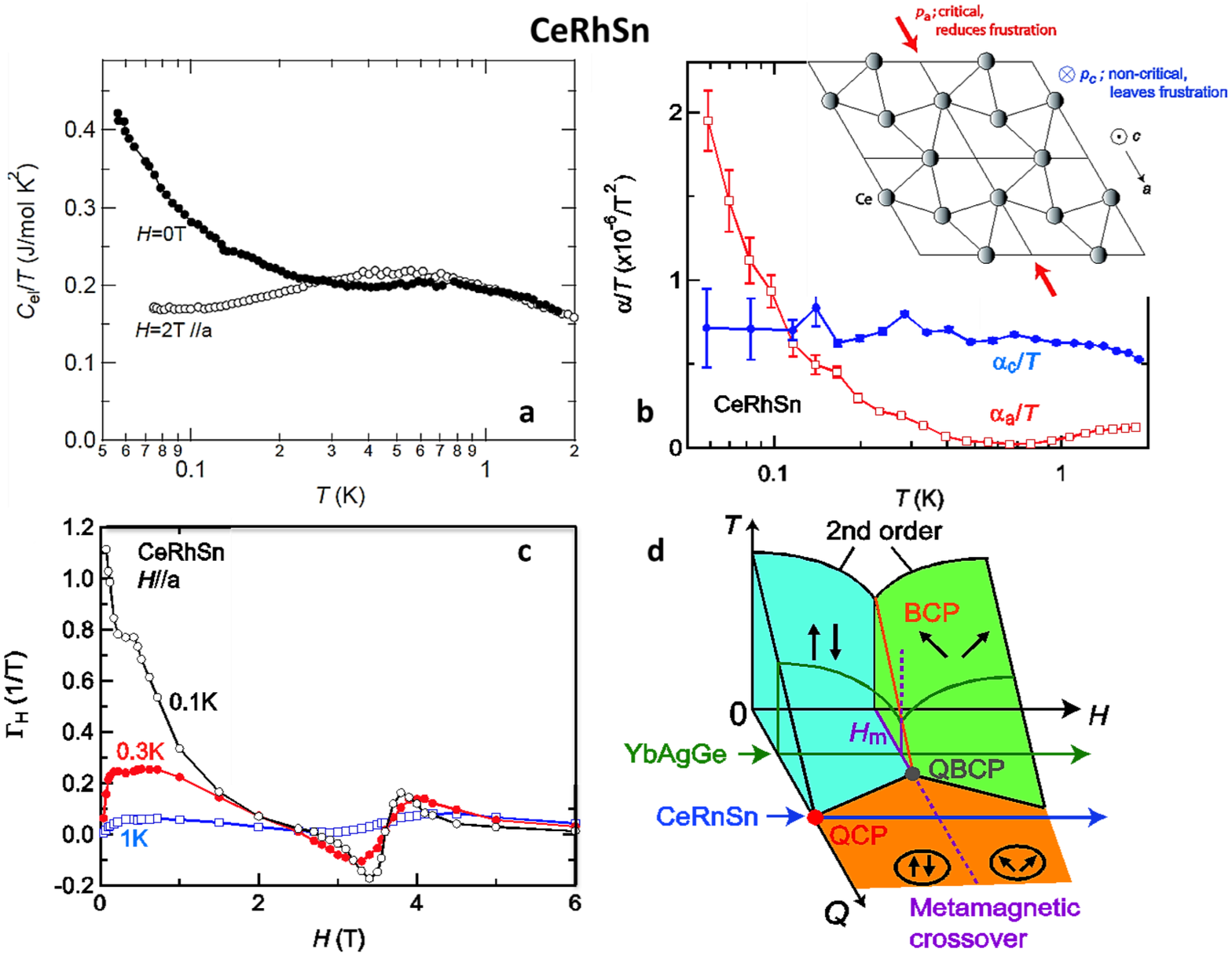}
	\caption{Evidence for a frustration induced QCP in hexagonal CeRhSn with distorted Kagome configuration of Ce atoms (cf. right inset)~\cite{TokiwaCeRhSn}. Specific heat coefficient (a) and linear thermal expansion coefficient (b) vs. $T$ (on log scale). Magnetic Gr\"uneisen ratio vs field (c). Schematic 3D temperature vs. magnetic field vs. frustration phase diagram, where $Q$ indicates the strength of quantum fluctuations induced by geometrical frustration. Compared to Fig. 3, it includes a first-order metamagnetic transition between differing AF$_S$ states (in blue and green). The evolution from a finite temperature bicritical point through quantum bicriticality towards a zero-field QCP and metamagnetic crossover within the P$_S$ (cf. Fig. 3) spin-liquid regime (in orange) is illustrated. The green and blue arrows indicate the positions of YbAgGe and CeRhSn in the phase diagram, respectively.}
\end{figure}

We now turn to quantum critical behavior in isostructural CeRhSn. Despite a large single ion Kondo scale of order 100~K which would indicate a tendency towards valence fluctuating behavior, the magnetic susceptibility is highly anisotropic and does not saturate upon cooling to low temperatures~\cite{Kim}. The absence of long-range order has been confirmed down to 50 mK by $\mu$SR~\cite{Schenk}. As shown in Fig.~9a, the heat capacity coefficient displays an unusual increase below 0.5~K, which is suppressed by the application of a moderate magnetic field. Indeed the Gr\"uneisen ratio of (volume) thermal expansion to specific heat, as well as the magnetic Gr\"uneisen ratio at low fields both display a power-law divergence indicating a zero-field QCP~\cite{TokiwaCeRhSn}. Moreover, the linear thermal expansion coefficient displays a pronounced anisotropy at low temperatures. While anisotropic thermal expansion behavior in general is not uncommon for non-cubic materials, it is unique to find NFL behavior along one direction (cf. the divergence of $\alpha_a/T$ in Fig. 9b) and ordinary FL behavior along another (here along the c-axis). Generally, thermal expansion is given by the sum of a background and a singular NFL contribution. Both are determined by the uniaxial pressure dependences of the respective entropy contributions. Thus, the striking anisotropy indicates that the quantum critical contribution to entropy has little c-axis pressure dependence. As sketched in Fig. 9b, this is expected for criticality arising from geometrical frustration, which is unaffected by c-axis pressure~\cite{TokiwaCeRhSn}. In the 3D phase diagram (Fig. 9d), this places CeRhSn at a zero-field QCP which is driven by strong quantum fluctuations due to geometrical frustration. Therefore some reminiscence of local-moment metamagnetism is expected. While the magnetic Gr\"uneisen ratio for fields along the c-axis monotonically decreases with increasing field, it shows a characteristic zero crossing near 3.5 T for the field along the a axis~(Fig.~9c)~\cite{TokiwaCeRhSn}. Along this direction the low-field low-$T$ magnetic susceptibility is about 30 times smaller compared to the respective data along the c-direction. Such behavior is different to itinerant metamagnetism, which is most pronounced along the direction with larger susceptibility. Rather it resembles local-moment metamagnetism. However, in contrast to ordinary spin-flop transitions, metamagnetism in CeRhSn does not occur as a first-order transition. Related to the fact that there are no ordered phases at the low- and high-field sides of the critical field, there is no first-order transition but only a weak crossover. This behavior could be ascribed to the strong geometrical frustration which suppresses long-range ordering. Furthermore, the involved moments are tiny as evidenced by the small entropy contribution of order $0.02 R\ln 2$. Kondo screening and quantum fluctuations are held responsible for the smallness of the moments. Altogether the thermodynamic properties suggest a novel type of QCP and metallic spin liquid state in CeRhSn~\cite{TokiwaCeRhSn}.

\section{Cooling through quantum criticality}

\begin{figure}\centering
	\includegraphics[width=\textwidth]{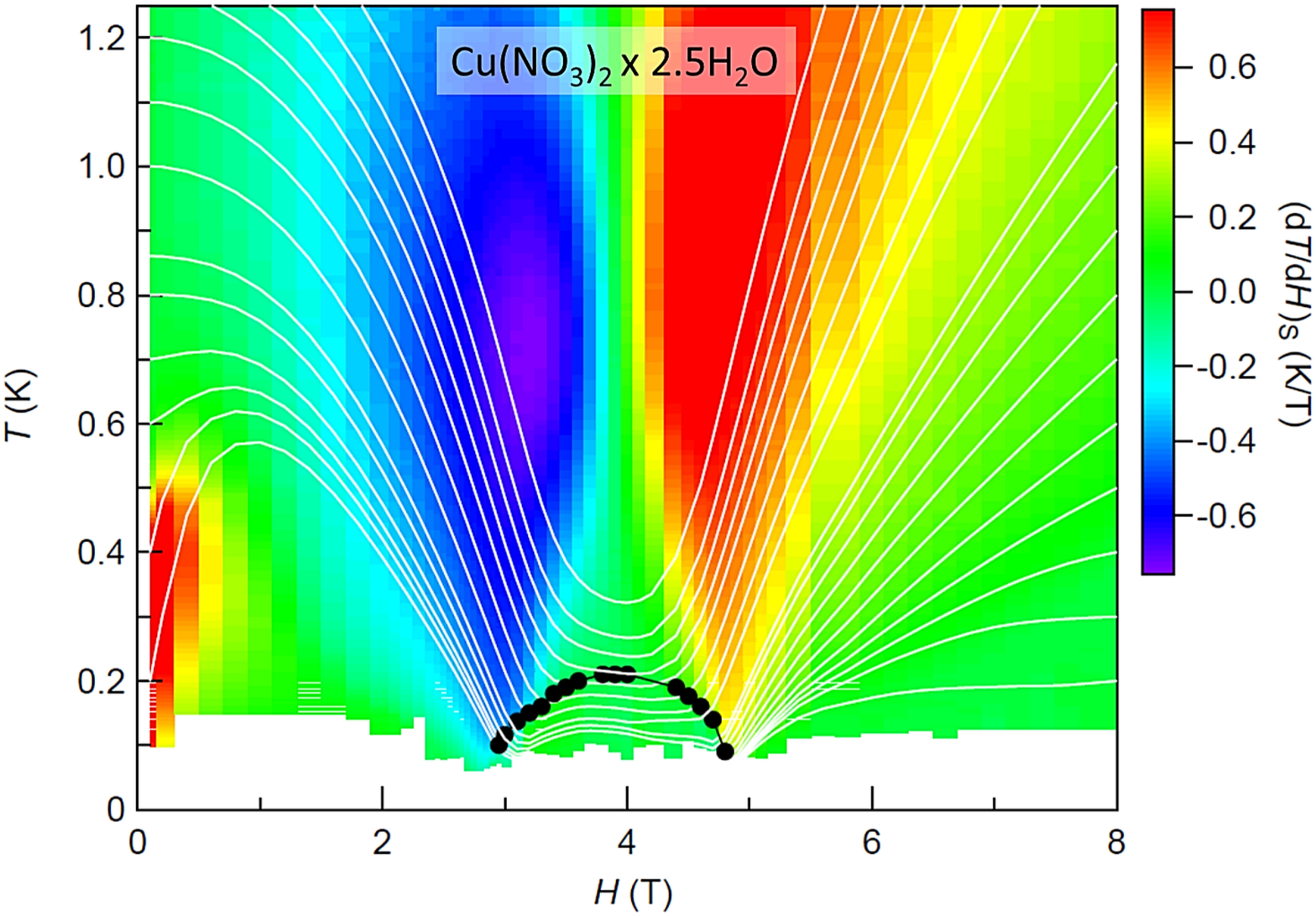}
	\caption{Adiabatic MCE effect for the $S=1/2$ coupled dimer system Cu(NO$_3$)$_2 \times 2.5$~H$_2$O~\cite{Diederix}. The white lines display adiabatic temperature traces, calculated by integration of the measured $(dT/dH)_S$ (cf. color coding) data~\cite{Tokiwa_unpublished}. The enhancement of the MCE at very low $T$ and low $H$ results from a small concentration of Cu$^{2+}$ impurities.}
\end{figure}

As discussed in Chapter 2, a field-induced QCP generically results in a divergence and sign change of the magnetic Gr\"uneisen ratio $\Gamma_H=T^{-1} (dT/dH)_S$. For illustration, Fig.~10 displays data for insulating copper-nitrate (CN), which is a well known dimerized 1D quantum magnet~\cite{Diederix,Tennant}. In CN an applied magnetic field induces spin-flop ordering between two critical fields, leading to two field induced QCPs. From the measured magnetic Gr\"uneisen ratio the adiabatic temperature traces are obtained by integration (cf. white lines in Fig. 10). They indicate a drastic cooling in the vicinity of the two QCPs, which has been known experimentally and modeled theoretically since several decades~\cite{Diederix}.

More recently it was recognized that compared to the adiabatic cooling with paramagnetic salts, cooling using quantum critical matter could have significant advantages~\cite{Wolf}. The latter allows in principle a much higher density of magnetic ions compared to paramagnetic salts which need to be very diluted for reducing their ordering to very low $T$. Thus, the cooling power per unit-cell volume for quantum critical systems could be considerably larger. Quantum critical materials, due to the abundance of low-energy excitations, could be cooled in principle to arbitrarily small minimal temperatures, keeping the cooling power large~\cite{Wolf}. By contrast, cooling with diluted paramagnetic salts is limited by the residual interaction of the moments. Their heat capacity shows a narrow Schottky type anomaly and highly efficient cooling is only reached in the temperature interval around the specific heat maximum. Quantum critical systems display an excitation spectrum very different compared to simple paramagnets. Within the quantum critical regime, the specific heat coefficient often follows a logarithmic temperature dependence or weak power law. Thus, on raising the temperature the specific heat coefficient decreases far more gently compared to the Schottky-type behaviour $C/T\sim T^{-3}$ for paramagnets. This leads to much longer hold times at elevated temperatures compared to paramagnets~\cite{Wolf}. Furthermore, the efficiency, defined by the ratio of heat expelled at the high temperature stage in an adiabatic magnetization refrigerator in comparison to the heat removed during cooling, is significantly larger for quantum critical materials than for paramagnetic salts~\cite{Wolf}.

Compared to magnetic insulators, HF metals have a huge thermal conductivity at low temperatures, which makes applications at very low temperatures significantly easier. However, in the regime close to the QCP typically only a very small amount of the entropy is involved in these materials due to the onset of Kondo screening as the temperature is lowered below $T_{\rm K}$. For Ge-doped YbRh$_2$Si$_2$ with $T_{\rm K}\approx 20$~K, for example, only about 10\% (2\%) of $R\ln 2$ is available below 1~K (0.1~K)~\cite{GegenwartYRS05}, which is too small for practical applications. Ideal for involving a significant amount of entropy in the cooling process would be a HF metal with small $T_{\rm K}$, being located close to a QCP at very low magnetic fields. Such constraints are very hard to fulfill, since Kondo lattices with low $T_{\rm K}\sim 1$~K typically display stable long range magnetic ordering with a large critical field of order 10~T or more. Cubic YbCo$_2$Zn$_{20}$ may be a promising candidate system, since it exhibits among all known paramagnetic HF metals the lowest Kondo temperature of only $T_{\rm K} = 1.5$~K~\cite{Torikachvili}. Its unique behaviour is related to its crystal structure which consists of Zn cages surrounding the Yb atoms. These cages lead to a very weak hybridization between the Yb 4f and Co 3d electrons. This effectively reduces $J$, which enters both $T_{\rm K}$ and $T_{\rm RKKY}$. Consequently, YbCo$_2$Zn$_{20}$  has an extremely large Sommerfeld coefficient, $\gamma= 8$~ J/mol K$^2$~\cite{Takeuchi}. Due to the small Kondo scale the HF state in this material can be effectively suppressed by fields of order several Tesla. YbCo$_2$Zn$_{20}$ can be driven towards a QCP by suitable doping and is a promising candidate for adiabatic cooling applications~\cite{Tokiwatbp}.

\section{Summary and outlook}

The physical properties of HF metals based upon partially filled 4f or 5f shells are governed by the interplay of the on-site Kondo and the long-range RKKY exchange interaction. Generically, they have a low-lying characteristic temperature, called Kondo temperature, which is highly sensitive to changes of the f-conduction electron exchange. This gives rise to a strong pressure sensitivity of the Kondo temperature and relatedly a highly enhanced Gr\"uneisen parameter. Similar as the enhanced Sommerfeld coefficient this is an important experimental indication of HF behavior. 

Furthermore, the Gr\"uneisen ratio of thermal expansion to specific heat and respectively the magnetic Gr\"uneisen parameter or adiabatic MCE are sensitive probes of QCPs. They have been used to characterize such instabilities in various material classes. Examples include insulating low-dimensional quantum spin chains~\cite{Wolf,Lorenz,Weickert12}, itinerant ferromagnets~\cite{Huy,Baier,Pfleiderer2007} or iron-pnictides~\cite{Meingast}.

In this review, we have concentrated on HF metals. The stronger than logarithmic divergence of the Gr\"uneisen ratio upon cooling in the quantum critical regime is a clear-cut criterion for the simple experimental identification of QCPs. It allows to distinguish smeared quantum phase transitions from generic QCPs. Furthermore a scaling analysis of the Gr\"uneisen ratio provides important information on the nature of the underlying critical fluctuations. We have categorized different materials by studies of the Gr\"uneisen parameter divergence. While one class shows criticality in accordance with the generic Hertz-Millis theory we have presented other examples, for which more advanced scenarios are currently discussed. Possibly the divergence of the magnetic Gr\"uneisen parameter can be used for cooling applications at mK temperatures.

The analysis of quantum critical behavior in the thermal expansion and the Gr\"uneisen ratio discussed in this article assumes a weak perturbation of criticality by the lattice degrees of freedom. This means that the latter do not modify the critical behavior. While this assumption is justified in many cases it breaks down for a bilinear coupling between the strain tensor and the order parameter, which is allowed at end points of lines of first-order metamagnetic or Mott transitions and at electronic nematic transitions~\cite{Garst15}. The critical behavior is then governed by critical crystal elasticity, which can lead to dramatic effects such as the breakdown of Hooke's law or vanishing sound velocities~\cite{ZPG}. Up to now, there are no systematic investigations of the sound velocity and the elastic constants near QCPs, probably due the requirement of large single crystalline samples, which are not available for many cases. Very recently, a QCP in the series (Ca$_x$Sr$_{1-x}$)$_3$Rh$_4$Sn$_{13}$, related to a second-order structural instability, has been found~\cite{Grosche}. Remarkably, the phonon heat capacity contribution is enhanced by almost a factor five near the QCP, indicating strong phonon softening. It should be interesting to investigate the signature of such a structural QCP using the Gr\"uneisen parameter.

\section*{Acknowledgment}

Long-term collaboration with M.~Garst, C.~Geibel, R.~K\"uchler, Q.~Si, F.~Steglich, C.~Stingl, Y.~Tokiwa and F.~Weickert is gratefully acknowledged. I thank P.C.~Canfield, K.~Heuser, M.-S.~Kim, J.A.~Mydosh, R.S.~Perry, E.-W.~Scheidt, T.~Takabatake, O.~Tegus and O.~Trovarelli for proving high-quality single crystals of the different materials reviewed in this paper. In particular I am grateful to R.~K\"uchler and Y.~Tokiwa for their strong efforts in the development of miniaturized capacitive dilatometers and thermal platforms for alternating field adiabatic MCE measurements, to C.~Stingl for his critical reading of this manuscript and to K.~Grube and H.v.~L\"ohneysen for providing Fig. 2a. Part of this article was prepared at the Aspen Center for Physics, which is supported by National Science Foundation grant PHY-1066293.



\end{document}